\documentclass[12pt]{article}
\usepackage[utf8]{inputenc}
\usepackage{amsfonts}
\usepackage{amsmath}
\usepackage{amssymb}
\usepackage{indentfirst}
\usepackage{graphicx}
\usepackage[colorlinks]{hyperref}
\usepackage{subcaption}
\usepackage{cite}
\usepackage{array}
\usepackage{microtype}
\usepackage{soul}
\usepackage[normalem]{ulem}
\usepackage{cancel}
\usepackage{soul}

\numberwithin{equation}{section}

\makeatletter

\usepackage [labelfont=bf] {caption}

\begin{document}

\begin{titlepage}
\vspace{3cm}

\baselineskip=24pt

\begin{center}
\textbf{\LARGE{Phase transitions for charged planar solitons in AdS}}
\par\end{center}{\LARGE \par}

\begin{center}
	\vspace{1cm}
	\textbf{Andrés Anabalón}$^{\left(1\right)}$,
	\textbf{Patrick Concha}$^{\left(2\right)}$,
	\textbf{Julio Oliva}$^{\left(3\right)}$,
	\textbf{Constanza Quijada}$^{\left(3\right)}$,\\
	\textbf{Evelyn Rodríguez}$^{\left(2\right)}$,
	\small
	\\[5mm]
    $^{\left(1\right)}$\textit{Departamento de Ciencias, Facultad de Artes Liberales,}\\
    \textit{Universidad Adolfo Ibáñez, Av. Padre Hurtado 750, Viña del Mar, Chile.}
    \\[2mm]
    $^{\left(2\right)}$\textit{Departamento de Matemática y Física Aplicadas, }\\
	\textit{ Universidad Católica de la Santísima Concepción, }\\
\textit{ Alonso de Ribera 2850, Concepción, Chile.}
	\\[2mm]
	$^{\left(3\right)}$\textit{Departamento de Física, Universidad de Concepción,}\\
	\textit{Casilla 160-C, Concepción, Chile.}
	\\[5mm]
	\footnotesize
	\texttt{andres.anablon@uai.cl},
	\texttt{patrick.concha@ucsc.cl},
	\texttt{juoliva@udec.cl},
	\texttt{constanquijada@udec.cl},
	\texttt{erodriguez@ucsc.cl},
	\par\end{center}
\vskip 26pt
\begin{abstract}
In this work we study the phase transitions between the planar charged AdS black hole and the planar charged soliton. The planar soliton is obtained as a double analytic continuation of the charged black hole metric, which also involves analytically continuing the electric charge. We show that there are phase transitions between both solutions depending on the electric potential, magnetic flux and temperature. The analysis is carried out in the Grand-Canonical ensemble.
\end{abstract}
\end{titlepage}\newpage {} {\baselineskip=12pt }

\section{Introduction}

The AdS/CFT correspondence \cite{Maldacena:1997re, Gubser:1998bc, Witten:1998qj} plays an important
role in the study of phase transitions in strongly coupled field theories. As shown by
Hawking and Page in \cite{Hawking:1982dh}, there is a phase transition between spherical
AdS black holes and the thermal AdS spacetime. In the field theory side, the
Hawking-Page transition has been shown to capture many features of the
confinement-deconfinement phase transition \cite{Witten:1998zw}. However, when
the field theory is on $R^{n}\times S^{1}$, $S^{1}$ being the thermal
cycle, no phase transition is observed to take place. The black hole is
always dominant and thus, the theory is always deconfined.

As remarked in \cite{Horowitz:1998ha}, when the black hole is planar, the configuration is
actually devoided of non-trivial integration constants. Hence, there is a
black hole, but no well defined thermodynamics. This situation is solved by
simply considering one of the horizon coordinates to be compact. This induces
a topology change. The field theory is now defined on $R^{n-1}\times
S^{1}\times S^{1}$. The introduction of the new cycle implies the existence of
a negative Casimir energy. The space with the minimum energy in this case is
the AdS soliton \cite{Horowitz:1998ha}. Thus, it is natural to consider the mentioned
AdS soliton as the ground state of the theory. In this case, there is a phase
transition  \cite{Surya:2001vj}, analogous to the Hawking-Page transition, between
the planar AdS black hole and the AdS soliton. As expected, the phase
transition is controlled by the dimensionless quantity $\beta/\beta^{\prime}$
where $\beta$ is the period of the thermal cycle and $\beta^{\prime}$ is the
period of the spacelike cycle. This has been rephrased as the fact that large
cold black holes, corresponding to the deconfinement phase, are stable, while
small hot soliton are stable corresponding to the confined phase.

A phase transition at zero temperature was presented in \cite{Banerjee:2007by} and
subsequently treated in \cite{Nishioka:2009zj, Horowitz:2010jq} for a particular critical value of the
chemical potential $\mu$. In particular, a first order phase transition occurs
between the neutral AdS soliton and the AdS charged black hole. The existence of two
intensive quantities makes possible to resolve the phase transition, finding that
it extends from a point to a line when $\beta/\beta^{\prime}$ and $\mu
/\beta^{\prime}$ are varied. An interesting extension of this development
considering Ricci-flat black holes and deformed AdS soliton has
been studied in the context of Gauss-Bonnet and dilaton gravity \cite{Cai:2007wz}. Although the black hole solution and AdS soliton are deformed by the Gauss-Bonnet term, it does not affect the Hawking-Page phase transition. Nevertheless, in dilaton gravity, the high temperature phase can be dominated either by black holes or deformed AdS solitons depending on the dilaton coupling constant. More recently, in \cite{Anabalon:2019tcy} it was also explored the phase diagram of four-dimensional neutral hairy planar black holes and solitons in AdS spacetime. In particular, it was presented how the hair affects the phase transition, showing that there exist first order phase transitions between the hairy black hole and the AdS soliton with the same boundary conditions.  

In this paper, we shall investigate the phase transitions that may occur between the four-dimensional charged planar black holes and charged planar AdS solitons. For this purpose, we consider the planar soliton as the thermal background. The aforementioned soliton is obtained as a double analytic continuation of the metric of the planar charged black hole, which also involves analytically continuing the electric charge. Thus, the planar soliton can be interpreted as a charged soliton solution as we are providing it with a gauge potential with a non-trivial profile, generalizing the work done in \cite{Surya:2001vj,Banerjee:2007by}.
We show that, in four dimensions, there are phase transitions between both solutions depending on the electric potential $\phi$, magnetic flux $\Phi$ and temperature $\beta^{-1}$. Interestingly, we show that there is a critical value for each of the aforesaid quantities for which the phase transitions can take place. Naturally, for a vanishing magnetic flux we recover the results presented in \cite{Banerjee:2007by} in which a phase transition at zero temperature occurs for a particular value of the electric potential.

The novelty of our result appears when the magnetic flux is switched on. Indeed, a phase transition also occurs at zero temperature between the charged AdS soliton and the AdS uncharged black hole for a particular value of the magnetic flux. Interestingly, in presence of both electric potential and magnetic flux, we show that there are phase transitions at zero temperature for a family of values of both quantities. In particular, we obtain a first order phase transition curve at zero temperature depending on the magnetic flux and electric potential. The phase transition surface for arbitrary values of temperature is also obtained. The surface defines a dome with respect to the axes $\phi$, $\Phi$ and $\beta^{-1}$, below which the charged soliton phase dominates while above it the charged black hole phase dominates.

The present paper is organized as follows: In Section 2, we present the general setup for a charged planar AdS black hole and the construction of the charged planar soliton. In Section 3, we compute the free energy considering the background subtraction method supplemented with the local counterterms \cite{Balasubramanian:1999re,Emparan:1999pm}. Then, we study in detail the phase structure in the Grand-Canonical ensemble and show that there exist phase transitions between the charged planar black hole and the charged planar soliton, which depend on the electric potential, magnetic flux and temperature. Section 4 is devoted to some concluding remarks and possible future developments.

\section{Planar AdS solitons and AdS black holes with charge}

In this section, we give the general setup of the charged AdS black hole with Ricci-flat horizon and the construction of the charged AdS soliton first presented in \cite{Anabalon:2021tua}. To start with, let us consider the $4$-dimensional Einstein-Maxwell-AdS theory, with action
\begin{equation}
I\left[  g_{\mu\nu},A_{\mu}\right]  =\int_{\mathcal{M}} d^{4}x\sqrt{-g}\left(
\frac{R}{2}-\frac{1}{8}F_{\mu\nu}F^{\mu\nu}+\frac{3}{\ell^{2}}\right)  \,,
\label{actEM}
\end{equation}
where $F(A)_{\mu\nu}=\partial_\mu A_\nu-\partial_\nu A_\mu$ and  $\ell$ is the AdS radius. The field equations are obtained by varying the action (\ref{actEM}) with
respect to the metric tensor $g_{\mu\nu}$ and the abelian gauge field $A_{\mu
}$
\begin{equation}
E_{\mu\nu}=R_{\mu\nu}-\frac{1}{2}g_{\mu\nu}R-\frac{3}{\ell^2}g_{\mu\nu}-\frac{1}{2}
T_{\mu\nu}^{A}=0\text{\thinspace},
\end{equation}
\begin{equation}
\partial_{\mu}\left(  F^{\mu\nu}\sqrt{-g}\right)  =0\,,
\end{equation}
where
\begin{equation}
T_{\mu\nu}^{A}=(F_{\mu\rho}F_{\nu}^{\text{ }\rho}-\frac{1}{4}g_{\mu\nu}
F^{\rho\sigma}F_{\rho\sigma})\,.
\end{equation}

\subsection{ Charged planar AdS black hole}
Let us consider the class of static metrics in $4$-dimensions
\begin{equation}
ds^{2}=-f_{b}(r)dt_{b}^{2}+\frac{dr^{2}}{\,f_{b}(r)}+\frac{r^2}{\ell^{2}}d\varphi_{b}^{2}+\frac{r^2}{\ell^2}dz^2\,, \label{staticmetric}
\end{equation}
By choosing this static ansatz, a solution of the previously introduced field equations is given by the following metric function 
\begin{equation}
f_{b}(r)=\frac{r^{2}}{l^{2}}-\frac{m}{r}+\frac{q^{2}}{{r^2}}\,,
\end{equation}
and
\begin{equation}
    A_{b}(r)=\left(\frac{2q}{r}-\frac{2q}{r_{+}}\right)dt_b + A_\varphi d\varphi_b
\end{equation}
where $m$ and $q$ are integration constants, $r_{+}$ is the largest root of the equation $f_{b}(r_{+})=0$ and $A_\varphi$ is a constant.
The coordinate $\varphi_{b}$ is periodic and has an arbitrary period $\eta_{b}$.

Upon Wick-rotating the metric \eqref{staticmetric}, we obtain the Euclidean black hole metric which is given by
\begin{equation}
ds^{2}=f_{b}(r)d\tau_{b}^{2}+\frac{dr^{2}}{\,f_{b}(r)}+\frac{r^2}{\ell^{2}}d\varphi_{b}^{2}+\frac{r^2}{\ell^2}dz^2\,,\label{euclideanBH}
\end{equation}

Note that the Wick rotation also affects the time-component of the gauge field, $A_{t}$. Regularity at $r=r_{+}$ demands that $\tau_{b}$ be identified with period $\beta_{b}=1/T$,
\begin{equation}
\beta_{b}=\frac{4\pi\ell^2r_{+}^{3}}{3r_{+}^{4}-q^2\ell^2}\ ,
\end{equation}
from where the temperature of the black hole reads
\begin{equation}
    T=\frac{1}{4\pi}\left(\frac{3r_{+}}{\ell^2}-\frac{q^2}{r_{+}^{3}}\right)\,.
\end{equation}

\subsection{ Charged planar soliton}

As in the neutral case, the $4$-dimensional charged AdS soliton is obtained as a double analytic continuation ($t\rightarrow i\varphi,$ $\varphi\rightarrow
it$) of the electrically charged black hole metric (\ref{staticmetric}). Then, the metric and the gauge field for the charged AdS soliton are written as

\begin{equation}
ds^{2}=-\frac{r^2}{\ell^2}dt_{s}^{2}+\frac{dr^{2}}{\,f_{s}(r)}+f_{s}(r)d\varphi_{s}^{2}+\frac{r^2}{\ell^2}dz^2\,, \label{planarsoliton}
\end{equation}
with 
\begin{equation}
f_{s}(r)=\frac{r^{2}}{l^{2}}-\frac{\mu}{r}-\frac{Q^{2}}{{r^2}}\,,
\end{equation}
and 
\begin{equation}
    A_{s}(r)=A_t dt_s + \left(\frac{2Q}{r}-\frac{2Q}{r_0}\right)d\varphi_{s}
\end{equation}
where $A_t$ is a constant, and $r_0$ is the largest root of $f_{s}(r_0)=0$. The point $r=r_0$ is a regular center provided the condition \eqref{etasfix} is fulfilled, therefore $r_0\leq r<\infty$. Note that in the above soliton solution the analytic continuation from the charged black hole also involves analytically continuing the charge \cite{Anabalon:2021tua}. As mentioned, the regularity of the spacetime at $r=r_{0}$ requires that coordinate $\varphi_{s}$ to be periodic, with the following period\footnote{The same procedure to construct soliton spacetimes from charged planar black holes, has been recently used also in \cite{Canfora:2021nca} in the context of $\mathcal{N}=4$ $SU(2)\times SU(2)$ gauged supergravity, giving rise to a new BPS spacetime.}
\begin{equation}\label{etasfix}
\eta_{s}=\frac{4\pi\ell^2r_{0}^{3}}{3r_{0}^{4}+Q^2\ell^2}\,.
\end{equation}
As discussed in \cite{Anabalon:2021tua}, the addition of a constant contribution of the gauge potential along $d\varphi$ ensures its regularity at $r=r_0$. Besides, we can define the net magnetic flux along the $z$ axis as follows,
\begin{equation}
    \Phi=-\int A_{\varphi}(r=\infty)d\varphi=\frac{2Q}{r_0}\eta_{s}\,.
\end{equation}
The Euclidean planar soliton is ($t_{s}\rightarrow-i\tau_{s}$)%
\begin{equation}
ds^{2}=\frac{r^2}{\ell^2}d\tau_{s}^{2}+\frac{dr^{2}}{\,f_{s}(r)}+f_{s}(r)d\varphi_{s}^{2}+\frac{r^2}{\ell^2}dz^2\,, \label{planarsoliton2}
\end{equation}
where $\tau_s$ is identified with period $\beta_s$, which is not restricted by any new regularity condition.

As we shall see, the matching conditions of the asymptotic regions of the charged black hole and charged soliton, make the period of $\tau_{s}$ equals to the period of
$\tau_{b}$ and the period of $\varphi_{b}$ equals to the period of $\varphi_{s}.$

\section{Phase transitions}
In this section, we analyze the phase transition between the planar charged AdS soliton and planar charged AdS black hole. The background subtraction method supplemented with the local counterterms \cite{Balasubramanian:1999re,Emparan:1999pm} is considered in order to explicitly study the phase transition, by computing Euclidean actions that can be interpreted as thermodynamics potential.

\subsection{Euclidean action for the charged black hole and charged soliton}
Let us now to describe the evaluation of the gravitational action for the planar charged black hole. As was previously discussed, we have to regularize the action, which for pure gravity is performed by supplementing it with the Gibbons-Hawking term and the counterterm $I_{\text{count}}$, such that
\begin{equation}
  I\left[  g_{\mu\nu},A_{\mu}\right]=I_{\text{bulk}}+I_{\text{GH}}+I_{\text{count}}\,,
\end{equation}
where $I_{\text{bulk}}$ is the term in \eqref{actEM}, and
\begin{align}
    I_{\text{GH}}=&\int_{\partial\mathcal{M}}d^{3}x\,\sqrt{-h}K\,,\\
    I_{\text{count}}=&-\int_{\partial\mathcal{M}
}d^{3}x\,\sqrt{-h}\left(\frac{2}{ \ell}+\frac{\ell}{2}\mathcal{R}\right)\,.
\end{align}
Here $K$ is the trace of the extrinsic curvature of the boundary, $h_{ij}$ is its induced metric and $\mathcal{R}$ is the Ricci scalar of boundary metric $h_{ij}$, which in this case vanishes. The Euclidean bulk  action reads
\begin{equation}
    I_{\text{bulk}}^E=\frac{L\eta_{b}\beta_{b}\rho^{3}}{\ell^4}-\frac{L\eta_{b}\beta_{b}m}{\ell^2}+\mathcal{O}\left(1/\rho\right)\,,
\end{equation}
where $\rho$ is the cut-off in the radial direction and $L=\int dz$. The contribution of the Gibbons-Hawking surface term and the local counterterm to the Euclidean black hole action can be respectively written as follows:
\begin{align}
     I_{\text{GH}}^E&=-\frac{3 L\eta_{b}\beta_{b}\rho^{3}}{\ell^4}+\frac{3 L\eta_{b}\beta_{b}m}{2\ell^2}+\mathcal{O}\left(1/\rho\right)\,,\\
     I_{\text{count}}^E&=\frac{2 L\eta_{b}\beta_{b}\rho^{3}}{\ell^4}-\frac{ L\eta_{b}\beta_{b}m}{\ell^2}+\mathcal{O}\left(1/\rho\right)\,.
\end{align}
  Then, the sum of all these terms gives
  \begin{equation}
  I_{b}^E= I_{\text{bulk}}^E+    I_{\text{GH}}^E+ I_{\text{count}}^E=-\frac{ L\eta_{b}\beta_{b}m}{2\ell^2}+\mathcal{O}\left(1/\rho\right)\,.\label{EuclBH}
  \end{equation}
  Notice that the Gibbons-Hawking term and the gravitational counterterm allow to cancel the divergent terms in the bulk action and also lead to a contribution to the $\mathcal{O}(1)$ terms.
Along the same lines, one can evaluate the Euclidean action of the planar AdS soliton, leading also to a finite expression which reads
\begin{equation}
  I_{s}^E= I_{\text{bulk}}^E+    I_{\text{GH}}^E+ I_{\text{count}}^E=-\frac{ L\eta_{s}\beta_{s}\mu}{2\ell^2}+\mathcal{O}\left(1/\rho\right)\,.\label{EuclSol}
  \end{equation}
\subsection{Free energy}
Let us now study the free energy and the phase transitions between the charged AdS soliton and the charged planar AdS black hole. To our purpose we calculate the action of the charged  black hole (\ref{euclideanBH})
of mass parameter $m$ and charge parameter
$q$, in the background of the soliton (\ref{planarsoliton2}).

In order to match the boundary geometry of both solutions, we have to consider the matching conditions at a finite radial cut-off $\rho$\, which includes:
\begin{align}
\sqrt{f_b(\rho)}\beta_{b} &
=\frac{\rho}{\ell}\beta
_{s}\,,\text{ \ \ }\\
\frac{\rho}{\ell}\,\eta_{b} &  =\sqrt{f_{s}\left(  \rho\right)  }\eta_{s}\,,
\end{align}
as well as the matching of metrics on such surface. Here, $\beta_{b}$, $\beta_{s}$, $\eta_{b}$ and $\eta_{s}$ are the periodicities of $\tau_{b}$, $\tau_{s}$, $\varphi_{b}$ and  $\varphi_{s}$, respectively.  In the limit $\rho\rightarrow\infty,$ we obtain that
\begin{equation}
\beta_{b}=\beta_{s}=\beta,\text{ \ \ \ }\eta_{b}=\eta_{s}=\eta\,. \label{matching}%
\end{equation}
Note that we also have to identify the electric potential of the black hole to that of the soliton and the magnetic flux of the soliton to that of the black hole. 

The Euclidean action of the black hole with respect to that of the soliton at the cut-off  $\rho$ is given by
\[
I^E(\rho)  =I_{b}^E (\rho) -I_{s}^E(\rho)  \,,
\]
where $I_{b}^E (\rho)$ and $I_{s}^E (\rho)$ are respectively given by \eqref{EuclBH} and \eqref{EuclSol}.
Considering the matching conditions (\ref{matching}) we find in the large
$\rho$ limit
\begin{equation}
I =\lim_{\rho\rightarrow\infty}I^E\left(  \rho\right)  =\frac{L \beta\eta}{2\ell^2}(\mu-m)\,. \label{euclideanaction}
\end{equation}
From the Euclidean action we can obtain the free energy $G=\beta^{-1}I$, which is given by
\begin{equation}
G=\frac{L \eta}{2\ell^2}(\mu-m)\,\,.
\end{equation}

 We can see that there is a change of sign of the free energy, depending on $\mu$ and $m$, that characterize the soliton and the black hole, respectively, which is a signal for a phase transitions. In order to work in the Grand-canonical ensemble we must express the free energy in terms of the fixed boundary conditions, namely the common temperature, value of $\eta$, chemical potential and magnetic flux. 

In consequence the energy reduces to
\begin{eqnarray}
    \tilde{G}&=&\left(1+\sqrt{1-X^2}\right)\left(1-2X^{2}+\sqrt{1-X^2}\right) \notag \\
    &&-\frac{1}{\Delta^{3}}\left(1+\sqrt{1+\Delta^2Y^2}\right)\left(1+2\Delta^2Y^{2}+\sqrt{1+\Delta^2Y^2}\right)\,,\label{free}
\end{eqnarray}
where, for simplicity, we have introduced the following rescaled quantities:
\begin{align}
    X&=\frac{\sqrt{3}}{4\pi \ell}\Phi\,,\\
    Y&=\frac{\sqrt{3}}{4\pi \ell}\eta\phi\,,\\
    \Delta&=\frac{\beta}{\eta}\,,\\
    \tilde{G}&=\frac{27\eta^2}{8\pi^3\ell^2L}G
\end{align}
In what follows we analyze the signs of $\tilde{G}$ as a function of the three independent rescaled thermodynamics variables $X,Y, \Delta$.
\subsection{Phase transitions}
From the expression of the rescaled free energy \eqref{free} we can observe that the phase transitions between the charged planar black hole and the charged planar soliton depend on the rescaled magnetic flux $X$, electric potential $Y$ and temperature $\Delta^{-1}$. In particular, we can see that 
\begin{align}
\tilde{G}&>0 \qquad \text{soliton phase dominates}\,,\\
\tilde{G}&<0 \qquad \text{black hole phase dominates}\,.
\end{align}

 To visualize the phase transitions we shall first plot the free energy \eqref{free}.
 
\begin{figure}[h]
    \begin{subfigure}{.5\textwidth}
    \centering
    \includegraphics[scale=0.38]{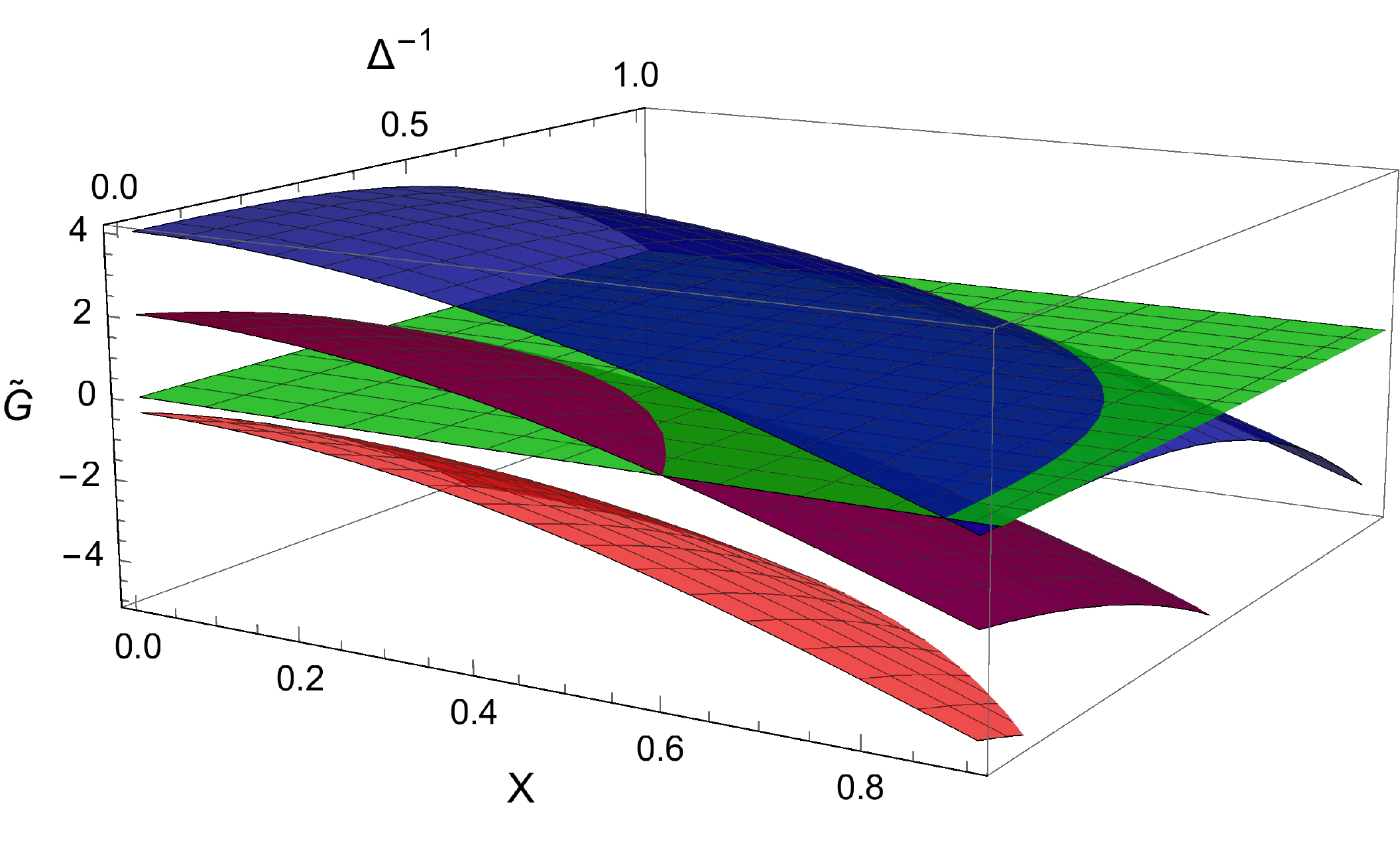}
    \caption{}
    \end{subfigure}
    \begin{subfigure}{.5\textwidth}
    \includegraphics[scale=0.39]{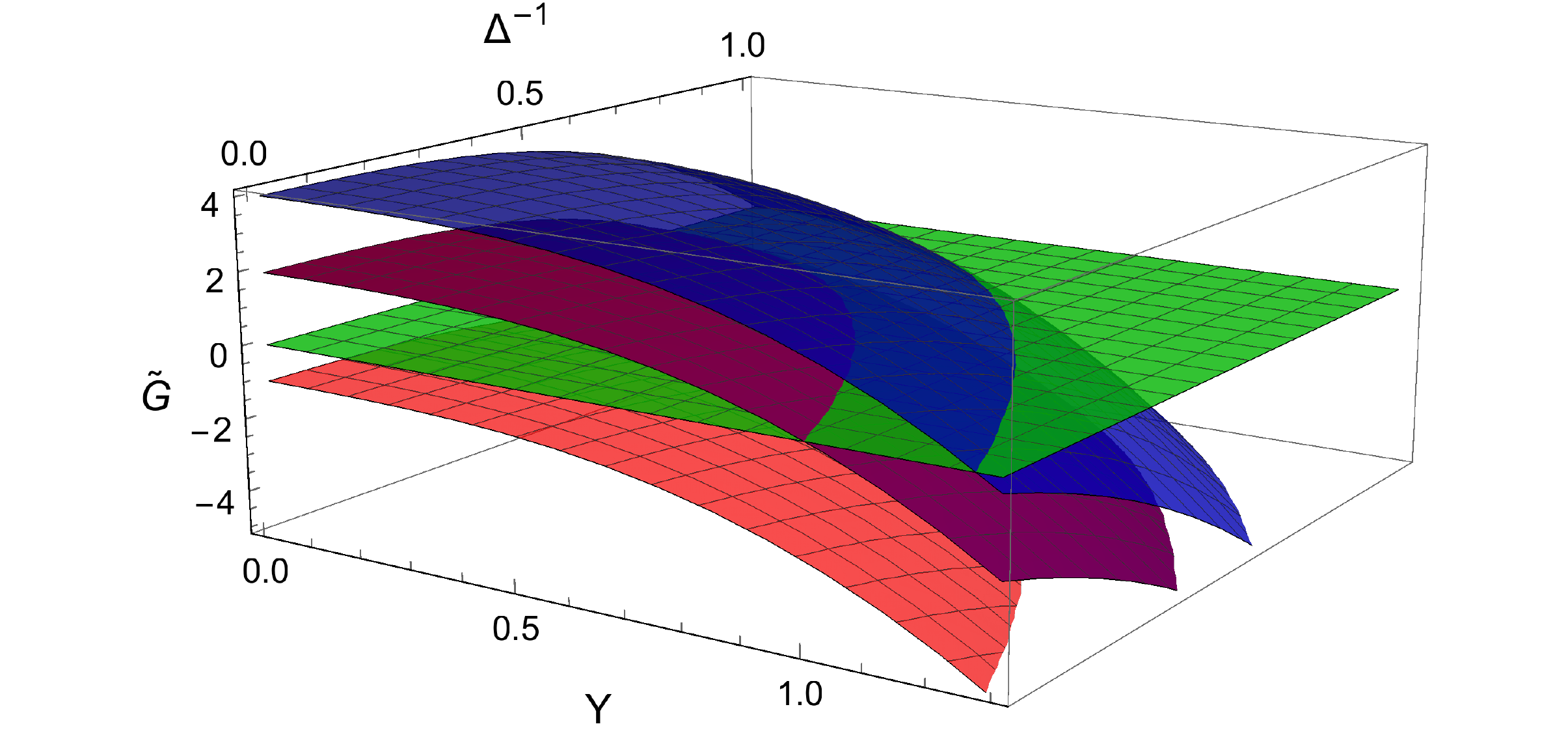}
    \caption{}
    \end{subfigure}
    \caption{(a) Free energy as function of temperature and magnetic flux for different values of electric potential. The values of the electric potential, from top to bottom, are $Y=0$, $Y=1.0$ and $Y=1.3$. 
    (b) Free energy as function of temperature and electric potential for different values of the magnetic flux. The values of the magnetic flux, from top to bottom, are $X=0$, $X=0.6$ and $X=1.0$.}
    \label{fig1}
\end{figure}

\begin{figure}[h!]
   \begin{subfigure}{.5\textwidth}
    \centering
    \includegraphics[scale=0.72]{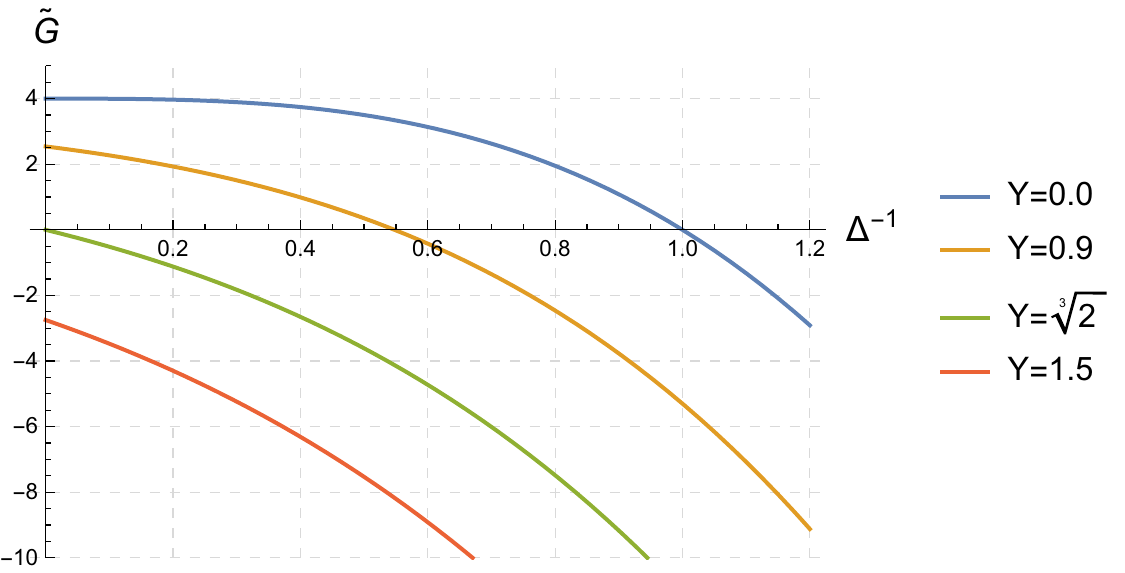}
    \caption{}
    \end{subfigure} \quad
    \begin{subfigure}{.5\textwidth}
    \includegraphics[scale=0.72]{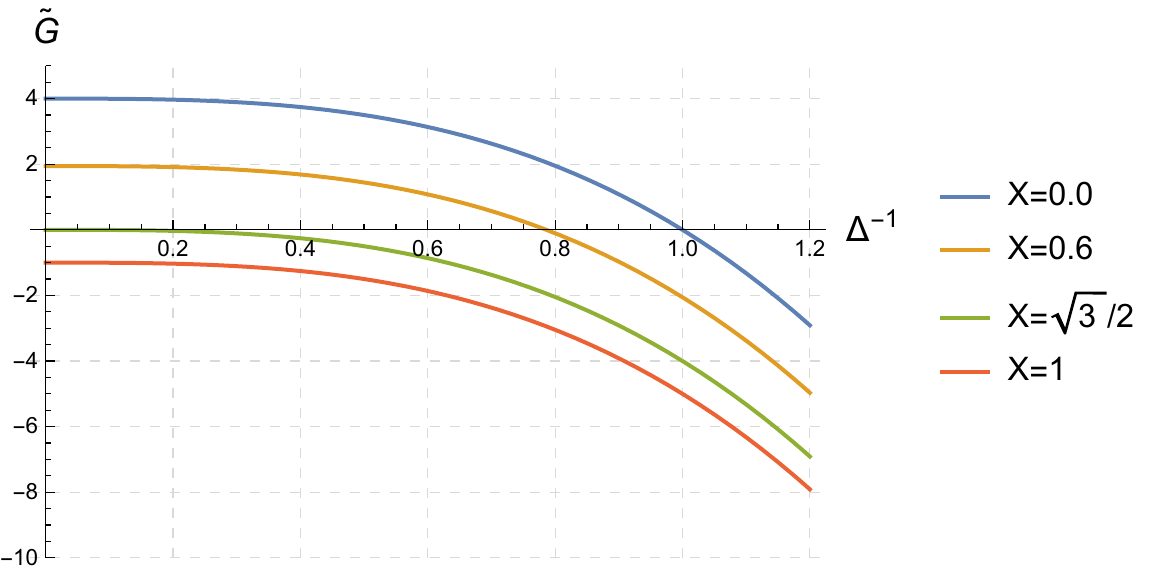}
    \caption{}
    \end{subfigure}
    \caption{(a) Free energy as function of temperature for vanishing magnetic flux $X=0$ and different values of electric potential.
    (b) Free energy as function of temperature for vanishing electric potential $Y=0$ and different values of magnetic flux.}
    \label{fig2}
\end{figure}

In Figure \ref{fig1} we can see that phase transitions occur for particular values of the electric potential $Y$, the magnetic flux $X$ and the temperature $\Delta^{-1}$. Indeed, from Figure \ref{fig2}, we can observe that phase transitions take place when the rescaled intensive thermodynamical quantities are bounded by the following critical values \footnote{The following bounds can be obtained analytically}
\begin{equation}
    X \leq \frac{\sqrt{3}}{2}, \quad Y \leq \sqrt[3]{2} \quad \mathrm{and} \quad  \Delta^{-1} \leq 1 \ . 
\end{equation}

One can see in Figure \ref{fig2} that for a temperature $\Delta^{-1}>1$, the (charged) black hole phase dominates independently of the values of the electric potential and magnetic flux. 

\begin{figure}[h!]
    \centering
    \includegraphics[scale=0.25]{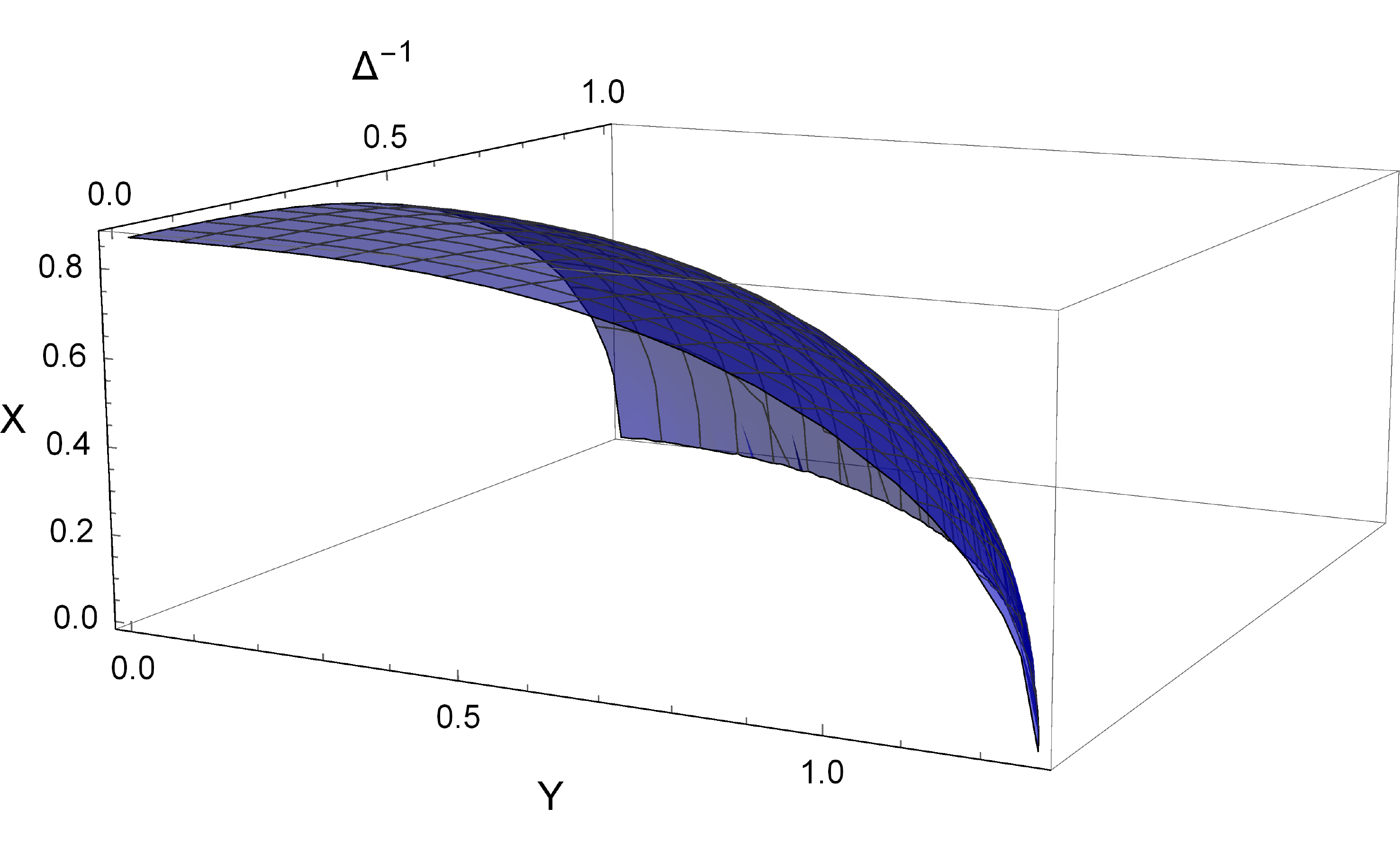}
    \qquad 
    \includegraphics[scale=0.25]{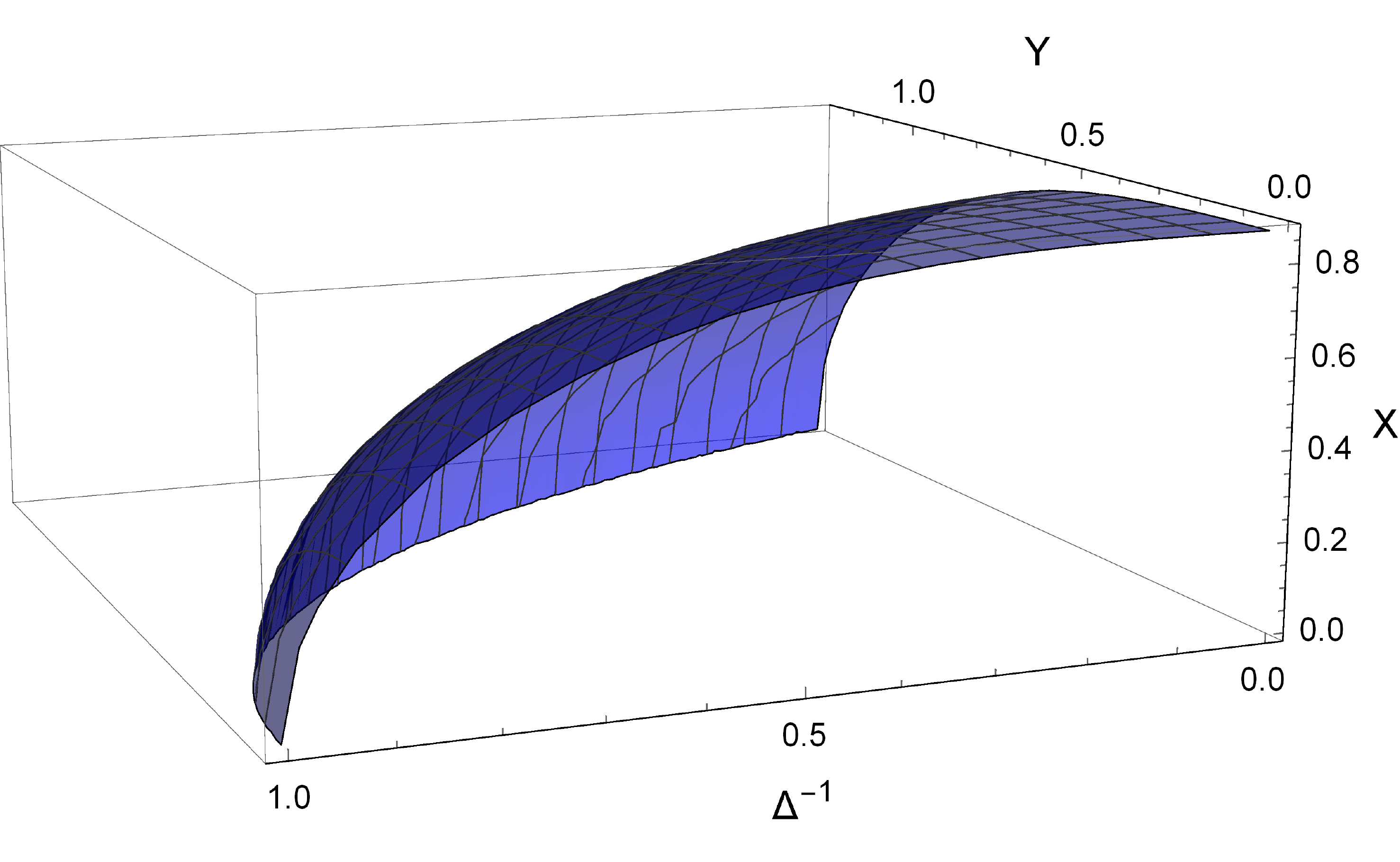} 
    \caption{Phase transitions surface between charged planar AdS black hole and charged planar soliton.}
    \label{fig3}
\end{figure} 

\begin{figure}[h!]
    \centering
    \includegraphics[scale=0.65]{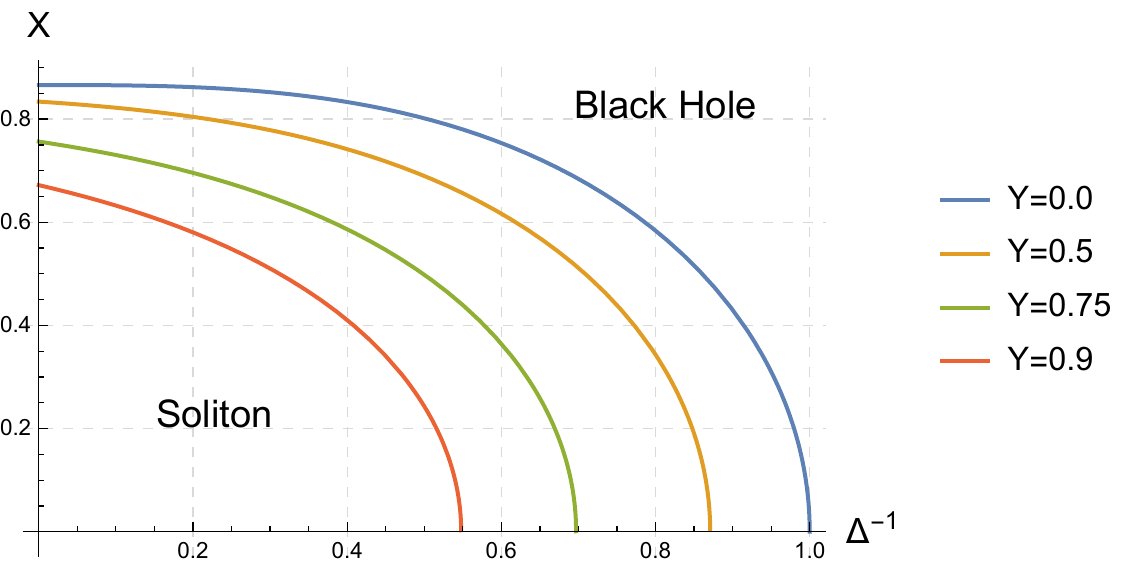} \qquad \quad
    \includegraphics[scale=0.65]{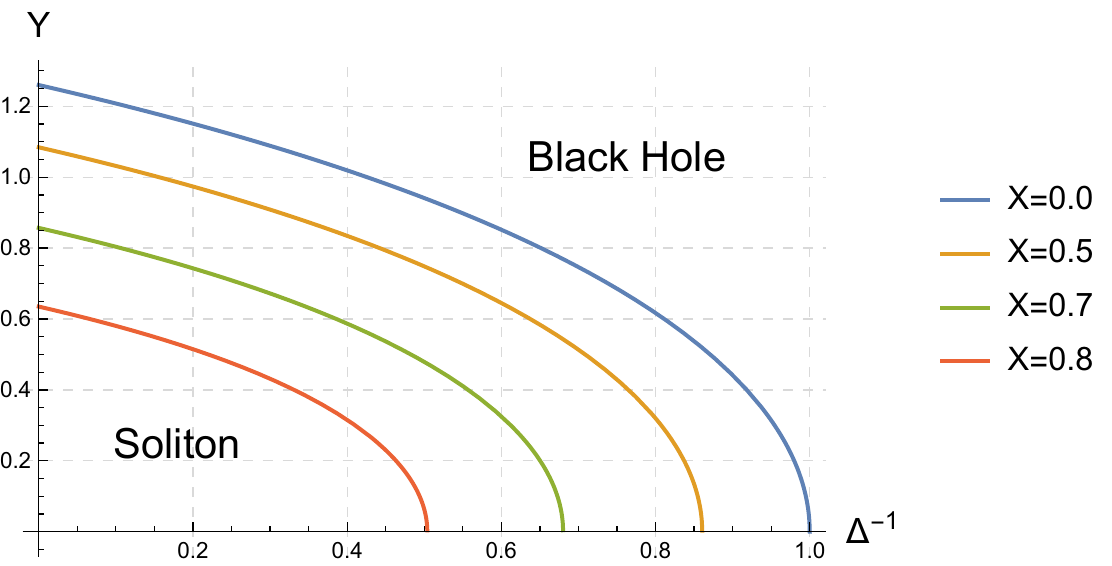} 
    \caption{Phase transitions curves between charged planar AdS black hole and charged planar soliton. On top of the curves a second order phase transition takes place, i.e, the two solutions coexist. Out of the curves, first order phase transitions occur. Below the curves the soliton phase dominates and above the curves the black hole phase dominates.}
    \label{fig4}
\end{figure}

The temperature, at which the phase transition occurs, changes with the electric potential and the magnetic flux as we can see in Figure \ref{fig3}. Above the phase transition surface the (charged) black hole phase is stable while below the surface the (charged) soliton solution is dominant. In the particular case of vanishing electric potential, the transition temperature occurs when
\begin{equation}
    \Delta^{-1}=\left[\frac{\left(1+\sqrt{1-X^2}\right)\left(1-2X^{2}+\sqrt{1-X^2}\right)}{4}\right]^{1/3}
\end{equation}\label{Trantemp}
which is depicted by the blue line in the plot $X$ vs $\Delta^{-1}$ of Figure \ref{fig4}. Remarkably, when $X=\sqrt{3}/2$, the phase transition temperature is zero. \\

From the graphic $Y$ vs $\Delta^{-1}$ of Figure \ref{fig4} we see that for $X=0$ we recover the results obtained in \cite{Banerjee:2007by} in which a first order phase transition occurs between the AdS charged black hole and the AdS soliton.

\begin{figure}[h!]
    \centering
    \includegraphics[scale=0.7]{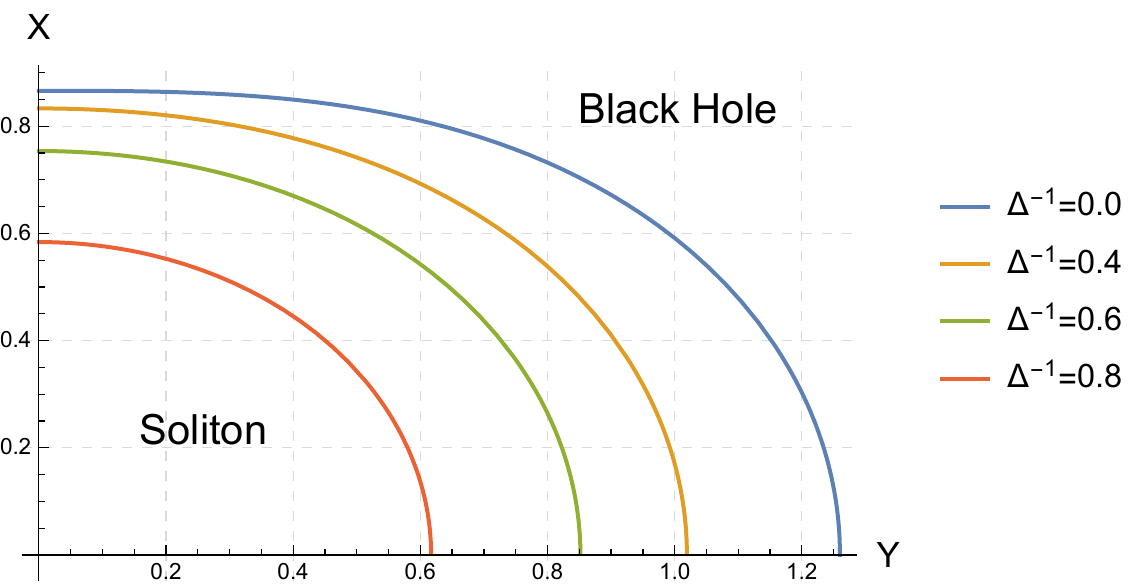} \qquad \quad
    \caption{Phase transitions curves for different values of temperature.}
    \label{fig5}
\end{figure}

The novelty appears when both, the electrical potential and the magnetic flux are switched on as the temperature approaches zero. From Figure \ref{fig5}, one can note that there is a family of values for the rescaled quantities in which a phase transition, between the charged planar black hole and the charged planar soliton, exists at zero temperature $\Delta^{-1}=0$. Indeed, the phase transition curve in this case is given by
\begin{equation}
 Y=\left[1+\sqrt{1-X^2}-\frac{1}{2}X^2\left(3+2\sqrt{1-X^2}\right)\right]^{1/3}
\end{equation} 

This particular feature, which can be seen as a generalization of the results obtained in \cite{Banerjee:2007by, Nishioka:2009zj, Horowitz:2010jq}, is quite different to the phase transition occuring between spherical AdS black holes and thermal AdS spacetime \cite{Hawking:1982dh} in which there is no phase transitions at zero temperature. Actually in such case there are no black holes below a critical temperature.

The following table summarizes the existing (see \cite{Banerjee:2007by}) and new phase transitions we have found, as well as the description of the transition loci in terms of the rescaled expressions for the electric potential $Y$, the magnetic flux $X$ and the temperature $\Delta^{-1}$:

\begin{table}[h]
\centering
\begin{tabular}{|m{6,6em}||m{3,6cm}|m{7,6cm}|}
\hline
 & Planar soliton & Charged planar soliton  \\ \hline\hline
Planar black hole & No phase transition & Phase transition for $X=\frac{\sqrt{3}}{2}$   \\ \hline
Charged planar black hole & Phase transition for $Y=\sqrt[3]{2}$ & Phase transition for 
$Y=\left[1+\sqrt{1-X^2}
-\frac{1}{2}X^2\left(3+2\sqrt{1-X^2}\right)\right]^{1/3}$   \\ \hline
\end{tabular}
\caption{Phase transitions at zero temperature}
\end{table}

\section{Discussion}

In this work we have extended the study of the phase transitions between the planar AdS black hole and planar soliton presented in \cite{Surya:2001vj}, by introducing charge to both black hole and soliton. The study of the free energy in four spacetime dimensions is approached considering the background subtraction method supplemented with the local counterterms. We have shown that there are phase transitions between the charged planar black holes and the charged planar soliton for particular values of the electrical potential $\phi$, magnetic flux $\Phi$ and the temperature $\beta^{-1}$. In particular, we have shown that the transition temperature depends on the electric potential and magnetic flux. For a vanishing magnetic flux, we have recovered the results obtained in \cite{Banerjee:2007by} in which there is a phase transition at zero temperature between the charged planar black hole and the planar soliton for a critical value of the electric potential.

Interestingly, in presence of magnetic flux,  we have shown that there is a phase transitions curve at zero temperature for specific values of both electric and magnetic quantities. In the limiting case in which the electric potential is zero, the phase transition at zero temperature occurs for a critical value of the magnetic flux. On the other hand, the critical temperature in which the phase transition takes place decreases as the electric potential $\phi$ and magnetic flux $\Phi$ is increased. 

It is tempting to argue that the phase transitions presented here could correspond to holographic  confinement/deconfinement transitions. It would be worth it understanding the deconfining phase related to the charged planar soliton in the condensed matter interpretation. 

\section{Acknowledgments}
This work was funded by the National Agency for Research and Development ANID - PAI grant No. 77190078 (P.C.), ANID - SIA grant No. SA77210097 (E.R.) and FONDECYT grants No. 1210635 (A.A.), 1181047 (A.A.), 1211077 (P.C.), 11220328 (P.C.), 1181047 (J.O.) and 11220486 (E.R.).  This work was partially supported by ANID Fellowships 21191868 (C.Q.) and the Research project Code DIREG$\_$09/2020 (P.C.) of the Universidad Católica de la Santisima Concepción, Chile. P.C. and E.R. would like to thank to the Dirección de Investigación and Vice-rectoría de Investigación of the Universidad Católica de la Santísima Concepción, Chile, for their constant support.



\bibliographystyle{fullsort.bst}
 
\bibliography{Phase_Transitions_VF}

\providecommand{\href}[2]{#2}\begingroup\raggedright\begin{thebibliography}{10}

\bibitem{Maldacena:1997re}
J.~M. Maldacena, ``{The Large N limit of superconformal field theories and
  supergravity},'' {\em Adv. Theor. Math. Phys.} {\bf 2} (1998) 231--252,
  \href{http://www.arXiv.org/abs/hep-th/9711200}{{\tt hep-th/9711200}}.

\bibitem{Gubser:1998bc}
S.~S. Gubser, I.~R. Klebanov, and A.~M. Polyakov, ``{Gauge theory correlators
  from noncritical string theory},'' {\em Phys. Lett. B} {\bf 428} (1998)
  105--114, \href{http://www.arXiv.org/abs/hep-th/9802109}{{\tt
  hep-th/9802109}}.

\bibitem{Witten:1998qj}
E.~Witten, ``{Anti-de Sitter space and holography},'' {\em Adv. Theor. Math.
  Phys.} {\bf 2} (1998) 253--291,
  \href{http://www.arXiv.org/abs/hep-th/9802150}{{\tt hep-th/9802150}}.

\bibitem{Hawking:1982dh}
S.~W. Hawking and D.~N. Page, ``{Thermodynamics of Black Holes in anti-De
  Sitter Space},'' {\em Commun. Math. Phys.} {\bf 87} (1983) 577.

\bibitem{Witten:1998zw}
E.~Witten, ``{Anti-de Sitter space, thermal phase transition, and confinement
  in gauge theories},'' {\em Adv. Theor. Math. Phys.} {\bf 2} (1998) 505--532,
  \href{http://www.arXiv.org/abs/hep-th/9803131}{{\tt hep-th/9803131}}.

\bibitem{Horowitz:1998ha}
G.~T. Horowitz and R.~C. Myers, ``{The AdS / CFT correspondence and a new
  positive energy conjecture for general relativity},'' {\em Phys. Rev. D} {\bf
  59} (1998) 026005, \href{http://www.arXiv.org/abs/hep-th/9808079}{{\tt
  hep-th/9808079}}.

\bibitem{Surya:2001vj}
S.~Surya, K.~Schleich, and D.~M. Witt, ``{Phase transitions for flat AdS black
  holes},'' {\em Phys. Rev. Lett.} {\bf 86} (2001) 5231--5234,
  \href{http://www.arXiv.org/abs/hep-th/0101134}{{\tt hep-th/0101134}}.

\bibitem{Banerjee:2007by}
N.~Banerjee and S.~Dutta, ``{Phase transition of electrically charged
  Ricci-flat black holes},'' {\em JHEP} {\bf 07} (2007) 047,
  \href{http://www.arXiv.org/abs/0705.2682}{{\tt 0705.2682}}.

\bibitem{Nishioka:2009zj}
T.~Nishioka, S.~Ryu, and T.~Takayanagi, ``{Holographic Superconductor/Insulator
  Transition at Zero Temperature},'' {\em JHEP} {\bf 03} (2010) 131,
  \href{http://www.arXiv.org/abs/0911.0962}{{\tt 0911.0962}}.

\bibitem{Horowitz:2010jq}
G.~T. Horowitz and B.~Way, ``{Complete Phase Diagrams for a Holographic
  Superconductor/Insulator System},'' {\em JHEP} {\bf 11} (2010) 011,
  \href{http://www.arXiv.org/abs/1007.3714}{{\tt 1007.3714}}.

\bibitem{Cai:2007wz}
R.-G. Cai, S.~P. Kim, and B.~Wang, ``{Ricci flat black holes and Hawking-Page
  phase transition in Gauss-Bonnet gravity and dilaton gravity},'' {\em Phys.
  Rev. D} {\bf 76} (2007) 024011,
  \href{http://www.arXiv.org/abs/0705.2469}{{\tt 0705.2469}}.

\bibitem{Anabalon:2019tcy}
A.~Anabalon, D.~Astefanesei, D.~Choque, and J.~D. Edelstein, ``{Phase
  transitions of neutral planar hairy AdS black holes},'' {\em JHEP} {\bf 07}
  (2020) 129, \href{http://www.arXiv.org/abs/1912.03318}{{\tt 1912.03318}}.

\bibitem{Balasubramanian:1999re}
V.~Balasubramanian and P.~Kraus, ``{A Stress tensor for Anti-de Sitter
  gravity},'' {\em Commun. Math. Phys.} {\bf 208} (1999) 413--428,
  \href{http://www.arXiv.org/abs/hep-th/9902121}{{\tt hep-th/9902121}}.

\bibitem{Emparan:1999pm}
R.~Emparan, C.~V. Johnson, and R.~C. Myers, ``{Surface terms as counterterms in
  the AdS / CFT correspondence},'' {\em Phys. Rev. D} {\bf 60} (1999) 104001,
  \href{http://www.arXiv.org/abs/hep-th/9903238}{{\tt hep-th/9903238}}.

\bibitem{Anabalon:2021tua}
A.~Anabalon and S.~F. Ross, ``{Supersymmetric solitons and a degeneracy of
  solutions in AdS/CFT},'' {\em JHEP} {\bf 07} (2021) 015,
  \href{http://www.arXiv.org/abs/2104.14572}{{\tt 2104.14572}}.

\bibitem{Canfora:2021nca}
F.~Canfora, J.~Oliva, and M.~Oyarzo, ``{New BPS solitons in $ \mathcal{N} $ = 4
  gauged supergravity and black holes in Einstein-Yang-Mills-dilaton theory},''
  {\em JHEP} {\bf 02} (2022) 057,
  \href{http://www.arXiv.org/abs/2111.11915}{{\tt 2111.11915}}.

\end{thebibliography}\endgroup

\end{document}